\documentclass[lettersize,journal]{IEEEtran}
\usepackage{amsmath,amsfonts}
\usepackage{algorithmic}
\usepackage{algorithm}
\usepackage{array}
\usepackage[caption=false,font=normalsize,labelfont=sf,textfont=sf]{subfig}
\usepackage{textcomp}
\usepackage{stfloats}
\usepackage{url}
\usepackage{verbatim}
\usepackage{graphicx}
\usepackage{cite}
\def\BibTeX{{\rm B\kern-.05em{\sc i\kern-.025em b}\kern-.08em
    T\kern-.1667em\lower.7ex\hbox{E}\kern-.125emX}}
\usepackage{balance}
\usepackage{url}
\usepackage{epstopdf}
\usepackage{amsmath}
\usepackage{amssymb}
\usepackage{amsthm}
\usepackage{booktabs}
\usepackage[table]{xcolor}
\usepackage{threeparttable} 
\usepackage{epsfig}

\usepackage{amsfonts}
\usepackage{multirow}
\usepackage[normalem]{ulem}
\useunder{\uline}{\ul}{}
\usepackage{color}
\usepackage{array}
\usepackage{acro}
\usepackage{listings}
\usepackage[underline=true]{pgf-umlsd}

\usepackage{setspace}

\DeclareRobustCommand{\acrodef}[2]{\DeclareAcronym{#1}{short=#1,long=#2}}
\acrodef{RRC}{Radio Resource Control}
\acrodef{ZMQ}{ZeroMQ}
\acrodef{LSTM}{Long Short-Term Memory}
\acrodef{LAL}{Listen-and-Learn}
\acrodef{SyAL}{Sync-and-Learn}
\acrodef{SoAL}{Source-and-Learn}

\acrodef{TS}{Technical Specifications}
\acrodef{NR}{New Radio}
\acrodef{URLLC}{Ultra-Reliable Low Latency Communications}
\acrodef{eMBB}{enhancing Mobile Broadband}
\acrodef{mMTC}{massive Machine Type Communications}
\acrodef{O-RAN}{Open Radio Access Network}
\acrodef{NAS}{Non-Access Stratum}
\acrodef{UE}{User Equipment}
\acrodef{MITM}{man-in-the-middle}
\acrodef{OTA}{over-the-air}
\acrodef{CN}{core network}
\acrodef{gNB}{gNodeB}
\acrodef{DoS}{Deny of Service}
\acrodef{BS}{Base Station}
\acrodef{RNTI}{Radio Network Temporary Identifier}
\acrodef{AR}{Augmented Reality}
\acrodef{IoT}{internet of things}
\acrodef{AI}{Artificial Intelligence}
\acrodef{SSD}{Systematic and Scalabile vulnerabilities
and unintended emergent behavior Detection Detection}
\acrodef{DCI}{Downlink Control Information}
\acrodef{CRC}{cyclic redundancy check}
\acrodef{ROC}{Receiver Operating Characteristics}
\acrodef{AUC}{Area Under the ROC Curve}
\acrodef{AKA}{Authentication and Key Agreement}
\acrodef{CDT}{cybersecurity digital twin}

\newcommand\algorithmicprocedure{\textbf{procedure}}
\makeatletter
\newcommand\PROCEDURE[3][default]{%
  \ALC@it
  \algorithmicprocedure\ \textsc{#2}(#3)%
  \ALC@com{#1}%
  \begin{ALC@prc}%
}
\newcommand\ENDPROCEDURE{%
  \end{ALC@prc}%

}

\newenvironment{ALC@prc}{\begin{ALC@g}}{\end{ALC@g}}
\makeatother

\begin{document}
%

\title{Radar Altimeter Redesign for Multi-Stage Interference Risk Mitigation in 5G and Beyond}
%
%
%



\author{Jarret Rock,~\IEEEmembership{Member,~IEEE,}
        Ying Wang,~\IEEEmembership{Member,~IEEE}

  \thanks{Jarret Rock and Ying Wang are with the School of Systems and Enterprises, Stevens Institute of Technology, Hoboken, USA (e-mail: jrock@stevens.edu; ywang6@stevens.edu).}
}

%
%

\markboth{ }%
{Shell \MakeLowercase{\textit{et al.}}: Bare Demo of IEEEtran.cls for IEEE Journals}
%



\maketitle

\begin{abstract}

The radar altimeter is installed on most 14 CFR Pt 25 category aircraft, which are applicable to passenger travel and represent most airline traffic. The radar altimeter system is highly accurate and reports the height above terrain. It plays a significant role in the take-off, approach, and landing phases of the applicable aircraft. In critical conditions, including reduced visibility, proximity to terrain, collision avoidance, and autoland procedures, the accuracy of radar altimeters is crucial to the safety of aircraft.

Effective January 2022, the permission for 5G deployment via the FCC, FAA, and other relevant parties generates safety concerns for the aviation industry, with multi-level impacts that include aircraft manufacturers, avionics equipment manufacturers, operators, air traffic controllers, certification authorities, and passengers. The 5G network operates in a frequency range that coincides with that previously dedicated to the radar altimeter system. The FAA has advocated the use of RF filters to purify the signals received by the radar altimeter. With shared frequency usage, considerations must be given to the interoperability of the systems. Furthermore, a hierarchy must be established regarding the criticality of the systems and which ones are susceptible versus offending. Inconsistent and inaccurate radar altimeter values communicated throughout the avionics architecture resulting from spurious 5G signals from cellular communication network use can be catastrophic.

This study aims to address the inappropriate behavior of the susceptible system that may cause essential safety concerns with unknown interoperability and operational impacts. We design and verify a strategic approach to mitigate the risks of potential airborne interference to a radar altimeter due to the coexistence of a 5G and future G signal, especially with the growing demand for the Space Air Ground Integrated Network (SAIGN). This study details a design change to a pre-existing radar altimeter system, and the process necessary to gain certification approval following this change is analyzed. We address the certification aspects from a TSO perspective resulting from changes made to a system post-certification. Artifacts, as defined in the FAA Project Specific Certification Plan template, including the Change Impact Analysis, Means of Compliance, and Test Plans, which are mandated by the certification authorities and requested by aircraft manufacturers and operators to ensure a level of compliance during the engineering cycle, have been adhered to.

\end{abstract}


%
\IEEEpeerreviewmaketitle

\section{Introduction}

The emergence of 5G network technology poses a safety concern for the aviation industry, particularly for the radar altimeter (also known as radio altimeter, rad alt, low-range altimeter, or RALT) system. This system is susceptible to RF interference due to 5G operating in close proximity to a dedicated aeronautical radio navigation frequency band. Of particular concern is the fact that 5G telecommunications operations occur in the range of 3.7 to 3.98 GHz \cite{RTCAInc2020AssessmentOperations}. Radar altimeters are precision systems that operate in the 4.2 to 4.4 GHz range and provide critical flight data to aircraft. An International Air Transport Association (IATA) and International Federation of Air Line Pilots’ Associations (IFALPA) document states that 'any failures or interruptions of these sensors can therefore lead to incidents with catastrophic outcomes, potentially resulting in multiple fatalities' \cite{IATA2020ProblemBand.}. At the time of writing this study, only preliminary suggestions existed for how to protect radar altimeters, and several manufacturers had not incorporated any such protection into their products. A proactive approach is needed to mitigate possible erroneous radar altimeter data resulting from unwanted interference while still maintaining airworthiness, and that is the focus of this study.


Radar altimeter systems are tasked with providing vertical height above terrain. This is an absolute value and is not influenced by factors such as atmospheric pressure and temperature. This is in contrast to baro-corrected altitude, which allows for pressure setting adjustments via a Kollsman Window as the aircraft transitions between air masses of varying pressure characteristics. Similarly, GPS data being used for landing applications is further refined by Wide Area Augmentation System (WAAS) and Receiver Autonomous Integrity Monitoring (RAIM) to guarantee a given accuracy \cite{Elrod2019The1}. Radar altimeters, however, typically have no compensation during their active mode and are thus dependent on a sanitized RF environment to guarantee accuracy. Given that radar altimeters are focal in providing data to the flight deck during landing phases when terrain separation is critical, safety concerns surround the reliability of this data in the presence of 5G emissions. Radar altimeters may be as accurate as $\pm3ft$ \cite{DepartmentofTransportationFederalAviationAdministrationAircraftCertificationService1966AIRBORNEALTIMETER}\cite{RTCAInc1974RTCA-DO-155Altimeters}.
\begin{figure}[!h]
    \centering
    \includegraphics[width=0.4\textwidth]{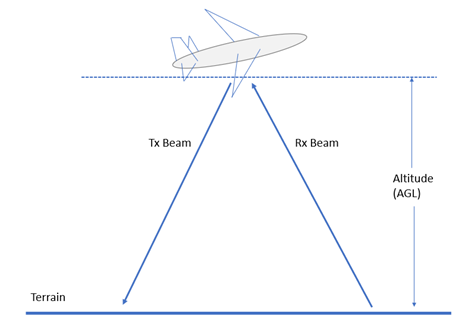}
    \caption{ Radar Altimeter Transmit and Receive Signal Propagation}
    \label{figure1}
\end{figure}

\begin{figure}[!h]
     \centering
     \includegraphics[width=0.5\textwidth]{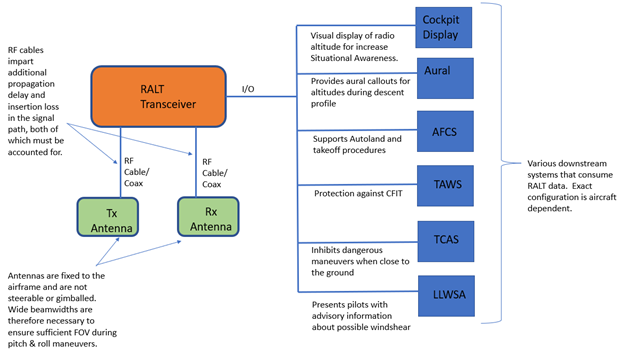}
     \caption{ System Block Diagram (Reproduced from: Honeywell, 2021)}
     \label{figure2}
 \end{figure}

Historically radar altimeters have been afforded frequency spacing from other systems hence protecting the integrity of the system.  According to DefenseNews, the C-Band is, “…relatively quiet…  For decades, this made the neighboring 4.2-4.4 GHz frequency a perfect place for the operation of radar altimeters…’\cite{Insinna2020TheSale}.  Due to the protected RF environment that radar altimeters have operated within, there is often no protection incorporated in the design of most legacy units against co-existent frequency usage.  There has previously not been any regulatory requirement to protect against 5G.  As stated in a Special Airworthiness Information Bulletin, “TSO-C87A does not provide criteria for compatibility with adjacent band operations, including potential impacts associated with wireless communications system deployments” \cite{DepartmentofTransportationFederalAviationAdministration2021SpecialAltimeters}. 
There is some collateral damage to consider.  The potential misleading data is not isolated to the radar altimeters in a monitor-only context.  Architecturally the radar altimeter provides altitude input for several other flight critical systems.  That is to say, an erroneous altimeter value is more than just a visual discrepancy on a pilot display, instead it can propagate this error into the Terrain Awareness Warning System (TAWS), the auto-land system (AFCS) and the Traffic Alert and Collision Avoidance System (TCAS) to name a few.  
Figure \ref{figure2} shows an example of a systems architecture with the rad alt feeding consumers of its data. 

In a civilian application, it should be noted that the radar altimeter is “…the only sensor that provides direct measurement of the clearance height of the aircraft over the terrain or other obstacles, and failures of these sensors can therefore lead to incidents with catastrophic results resulting in multiple fatalities” \cite{RTCAInc2020AssessmentOperations}.  Radar altimeters are featured on both civil and military aircraft installations and are poised to be a key system in autonomous flight vehicles.  An effective solution to protecting radar altimeter data can protect lives in aviation. 

This study will present one strategy that may be used by a radar altimeter manufacturer to reduce the susceptibility to 5G RF interference of currently fielded radar altimeter products.  Further, the study will highlight the engineering life cycle associated with making modifications to existing products that have been previously certified for aircraft use via a Technical Standard Order (TSO). The intellectual contribution from this study can be summarized as: 
\begin{itemize}
    \item Detailing the macro-process for revising an existing TSO’d product. Its benefits include a significant reduction in both Engineering effort and Certification approval time based on a delta certification strategy. 
    \item Creating the PSCP. Its benefits include a summary of the elements that create the PSCP and a definitive process for determining the Major or Minor classification of modifications to a sample radar altimeter. 
    \item Description of implementation of the solution. Its benefits include various means of demonstrating Compliance with a still evolving regulatory baseline. 
    \item Description of the equipment verification process.  Its benefits include exploring various verification means and their suitability to the regulations and standards that the equipment manufacturer is seeking compliance with. 
\end{itemize}

The rest of this paper is organized as follows. Section~\ref{background} discussed the background and related work. Section~\ref{system} describes the overview and system design of radar altimeter redesign and risk mitigation following the product cycle. Section~\ref{redesign} provides details I the radar altimeter redesign, followed by Section~\ref{verification} in verifying the design and certificate program for potential deployment. We discussed the current limitations and potentials in further collaborative work in Section~\ref{discussion}. Finally, the conclusion and future work is shown in Section~\ref{conclusion}. 

\section{Background and Related Work}\label{background}
The Radio Technical Commission for Aeronautics (RTCA) in conjunction with the Aerospace Vehicle Institute, has produced a paper titled Assessment of C-Band Mobile Telecommunications Interference Impact on Low Range Radar Altimeter Operations.  The RTCA Special Committee 239 formed a 5G Task Force in April 2020 to lead the research efforts that would produce the data featured in the paper.  It was determined by the aviation industry that there was a need to characterize the performance of fielded radar altimeters operating in the presence of RF interference from 5G networks in the band of concern.  Additionally, the aviation industry wanted to understand the risks of 5G and the potential impacts to continued safe aviation operations \cite{RTCAInc2020AssessmentOperations}.  Technical information was sourced from both the mobile industry as well as radar altimeter manufacturers.  One conclusion of the RTCA document is that the results presented within the “… report reveal a major risk that 5G telecommunications systems in the 3.7-3.98 GHz band will cause harmful interference to radar altimeters on all types of civil aircraft including commercial transport airplanes: business, regional and general aviation airplanes; and both transport and general aviation helicopters” \cite{RTCAInc2020AssessmentOperations}. 
The concern is beyond national, IATA and IFALPA have jointly released a document to address their concerns related to 5G.  Much of their findings are consistent with those of the RTCA. 
“Radar altimeters are deployed on tens of thousands of commercial and general aviation aircraft as well as helicopters worldwide.  The radar altimeter is one of the most critical components to an aircraft’s operations… Undetected failure of this sensor can therefore lead to catastrophic results…” \cite{IATA2020ProblemBand.}.

\begin{figure}[!h]
    \centering
    \includegraphics[width=0.5\textwidth]{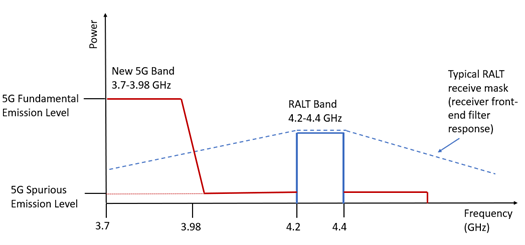}
    \caption{ Fundamental and Spurious 5G Emissions on Radar Altimeters }
    \label{figure3}
\end{figure}

In another study on the effects of EMI upon aircraft avionics, it is stated that, “The adverse impact of a 5G mobile handset cannot be understated.  This underscores the need to conduct comprehensive testing on EMI attack scenarios…”\cite{Solkin2021ElectromagneticSecurity_new}.  There is mention of the fact that mobile phones are a threat to aircraft navigational systems due to the high frequency and hence short wavelength of their signals.  Given that 5G phones operate in an even shorter wavelength, there is a concern that the signals are more intense as they are travelling shorter distances and will not have the opportunity to attenuate.  These short travelling signals can conflict aircraft navigation signals that are propagating in the immediate vicinity of the aircraft. \cite{}.  The article contends that there are at least two types of radiative interference from a cellular phone to be considered.  The necessary fundamental emissions required to operate the phone in conjunction with its base station (send and receive signals) but less predictable are the spurious emissions which are unwanted but may also fall within the C-Band.  The fundamental emissions occur outside the primary bandwidth of the radar altimeter hence the danger is blocking interference.  Spurious emissions are within the bandwidth of the radar altimeter and therefore either desensitize or lead to false determination of the radar altimeter.
.  The image in Figure \ref{figure3} reproduced from shows the effect of both the fundamental and spurious emissions on the radar altimeter band. 

\section{Overview of Product Life Cycle}\label{system}
The proposed method is to present a case study of a strategy to address 5G concerns with a rad alt product.  Focal to this is the addition of a bandpass filter at the recommendation of the FAA to mitigate interference\cite{DepartmentofTransportationFederalAviationAdministration2022FAA5G}.  The study addresses the certification process associated with revisions to an exiting TSO and re-integration into the aircraft.

\begin{figure}[!t]
    \centering
    \includegraphics[width=0.5\textwidth]{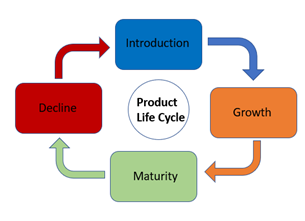}
\caption{Business Model and Product Life Cycle} 
    \label{figure4}
    \vspace{-15pt}
\end{figure}

The various stages of the product life cycle have long been established and have roots tracing to the economist Raymond Vernon in 1966 \cite{Osland1991OriginsConcept_new}.  This study assumes the four stages as shown in Figure   \ref{figure4} to support the radar altimeter product.  
After an assessment by the Product Management Team (PMT) that the radar altimeter product is in a profitable stage such as Maturity, it can justify the expenses associated with re-Engineering and re-Certification.

\begin{figure}[!h]
    \centering
    \includegraphics[width=0.4\textwidth]{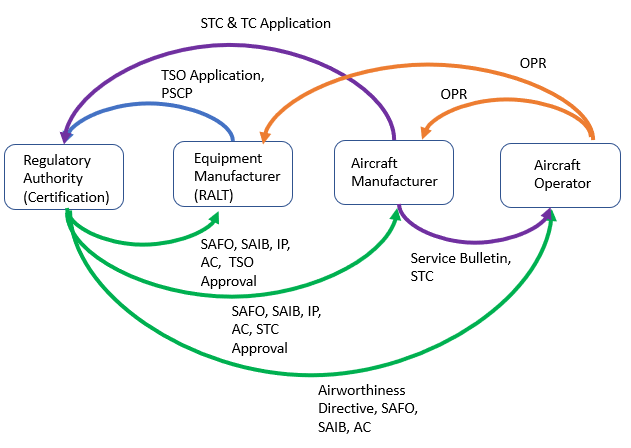}
    \caption{ Typical Communication Flows}
    \label{figure6}
\end{figure}

Communication during the product life cycle is key in ensuring that information is flowing appropriately between the various stakeholders of the TSO process.  Some of the key communication paths are provided in Figure \ref{figure6}.

“Airworthiness Directives are legally enforceable regulations issued by the FAA in accordance with 14 CFR part 39 to correct an unsafe condition in a product” \cite{DepartmentofTransportationFederalAviationAdministration20211439}.  As such, 
The FAA issued AD 2021-23-12 and AD 2021-13 for airplanes and helicopters respectively to address, “… a determination that radio altimeters cannot be relied upon to perform their intended function if they experience interference from wireless broadband operations in the 3.7-3.98 GHz frequency band (5G C-Band)”\cite{DepartmentofTransportationFederalAviationAdministration20211439}.  The ADs required operators to revise their AFM to incorporate limitations prohibiting certain operations that are radar altimeter dependent when there is a known presence of 5G signals as communicated by Notices to Airmen (NOTAM) \cite{DepartmentofTransportationFederalAviationAdministration2021RiskInterference1}  Similarly (Safety Alert for Operators) SAFO 21007 was issued to operators to inform that certain Instrument Approach Procedures are restricted if by NOTAM they are “affected by 5G C-Band interference, and prohibited by the ADs unless the operator has an FAA-approved AMOC” \cite{DepartmentofTransportationFederalAviationAdministration2021RiskInterference1}.  

\begin{figure}[!h]
    \centering
    \includegraphics[width=0.4\textwidth]{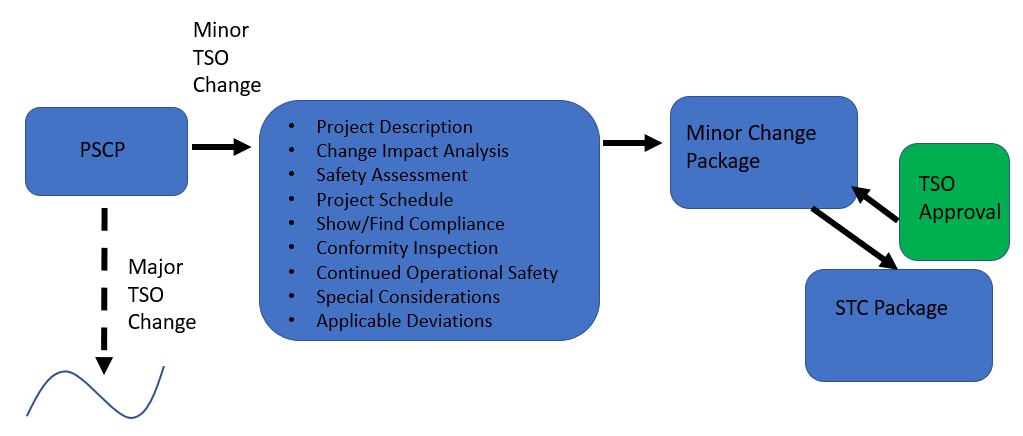}
    \caption{Sample Certification Package Flows}
    \label{figure7}
\end{figure}

With the various operational restrictions and the threat of invalidation of a TSO for a product that provides aircraft critical data, the longevity of the product is in question.  In an attempt to preserve the product, the TSO must be restored.  Following analyses by PMT, Engineering, Production, and Certification teams, a path forward is defined.  In the case of this particular product, it is assumed that its planned end-of-life is still distant and that there is a need to maintain the airworthiness of aircraft that utilize this radar altimeter.  A compliance strategy is then defined by the Certification team and Engineering is used to create, implement and verify that the proposed solution meets certification intent.  This plan to revise an existing product is communicated to the FAA by the TSO holder – the radar altimeter manufacturer, via a Project Specific Certification Plan (PSCP), refer to Figure \ref{figure6} and Figure \ref{figure7}.   
With the oversight of the Certification Authority, the manufacturer establishes a Certification Package Flow.  The Certification Authority confirms with the manufacturer that they can Find Compliance to the Affected Regulations via the artifacts generated from the Certification Package Flow.  Figure \ref{figure7} illustrates the Certification Package Flow relevant to this case study.

\section{Radar Altimeter Re-Design and Impacts Analysis }\label{redesign}
\begin{figure}[!h]
    \centering
    \includegraphics[width=0.5\textwidth]{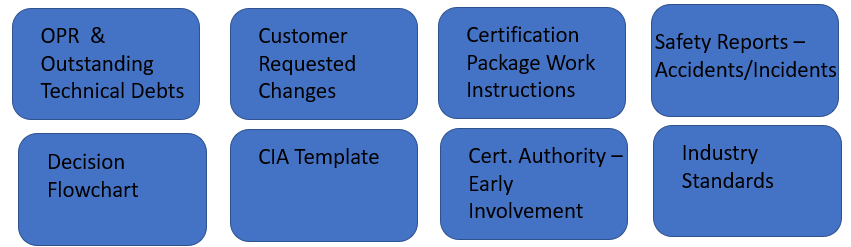}
    \caption{CIA Decision Inputs}
    \label{figure8}
\end{figure}

An output or deliverable from the PSCP is the Change Impact Analysis (CIA).  The CIA is used to identify the proposed changes for the equipment and to determine the impact to certification.  During the Engineering effort to create the CIA, Figure \ref{figure8} shows some of the details that need to be considered.

\textbf{Declaration of the Change}
In this case study the Declaration of the change is a ‘Minor Hardware Change'.  The distinction must be made as to whether the change is Minor or Major as there is a direct impact on the certification strategy or path, see Figure \ref{figure7}.  Major changes are subject to more development, verification and certification rigor.  The intended modification does not affect form, fit or intended function of the radar altimeter.  The hardware modification is merely incidental to the operation of the radar altimeter and requires no additional training for the operator.  It is classified as Minor, refer to the applicable regulatory definitions that classify these changes\cite{DepartmentofTransportationFederalAviationAdministratio202214Changes}.  This determination is guided by 14 CFR 21.619.  An assumption is that this modification does not require the full compliment of MOPS testing, thus it can be rationalized that the addition of the filter is not substantial enough to warrant a “…complete investigation to determine compliance with a TSO…” and therefore cannot be a Major change \cite{DepartmentofTransportationFederalAviationAdministratio202214Changes}. 

\textbf{Project Schedule}
The estimated project completion time is defined at the start of the project.  Schedules however are subject to factors such as finances, available Engineering resources to implement the change, Production lead times on components such as the filter, Production assembly times, laboratory availability to simulate the 5G environment, reconfiguring the aircraft for testing, FCC approval to conduct flight test as the radar altimeter is emissive and the FAA’s process time for TSO approvals. 

\textbf{Affected Part Number} 
The Affected PNs are identified at the time that the life cycle evaluation is conducted.  Given that there is a Hardware change associated with the modification, there is an evolution of the PN. 

\textbf{Modification Description}
The primary modification is to incorporate the changes as proposed by the FAA in their guidance on 5G \cite{DepartmentofTransportationFederalAviationAdministration2022FAA5G}.  This involves the installation of a bandpass filter on the receiver partition of the radar altimeter.  The filter will be installed prior to the RF input side of the rad alt.   
\begin{figure}[!h]
   \centering
   \includegraphics[width=0.5\textwidth]{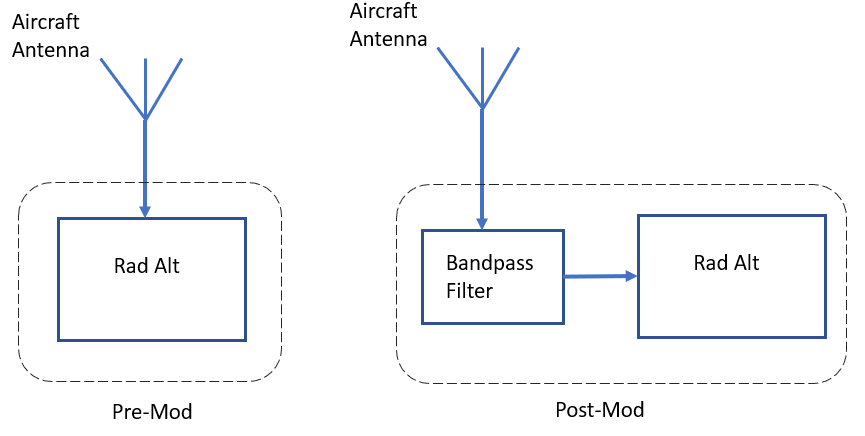}
   \caption{Pre and Post Modification of Rad Alt}
   \label{figure10}
\end{figure}

There are two Stopbands, one on either side of the desired Passband.  The theory is that only the intended frequency range of 4.2-4.4 GHz will pass through the filter.  The undesirable 5G frequencies of 3.7-3.98 GHz being restricted by the Stopbands.  The filter is proposed to also prevent spurious emissions from influencing the Bandpass. 

It must be determined if there is an insertion loss penalty due to the filter installation.  Compensations must account for unwanted signal attenuation either in software or a further hardware change.

It is typical that several hardware and/or software changes will be incorporated in a single TSO update.  This is due to the practice of reviewing Open Problem Reports (OPRs) and attempting to resolve them in bulk if there is another modification in process.  Figure \ref{figure8} Shows that an OPR review is required as part of the CIA process.  This case study does not include modifications beyond that required to incorporate the filter. 

\textbf{Affected Regulations/Requirements/Standards }
An analysis must be conducted of the applicable regulations.  The hypothetical radar altimeter in this study was previously granted TSO-C87 approval which is the FAA’s standard for radar altimeter performance.  The MOPS, DO-155 outline the minimum operational performance needed for a piece of equipment to meet TSO approval. The MOPS therefore are Affected Requirements and Standards and since issued under 14 CFR Part 37, Affected Regulations\cite{DepartmentofTransportationFederalAviationAdministration2017TechnicalProgram}.  Advisory Circular AC25-7D provides guidance for flight testing of transport category airplanes and thus is also an Affected Requirements document\cite{FederalAviationAdministration2018FlightAirplanes}.  There are self-imposed requirements, often to provide a competitive edge against other products within the market.  These self-imposed requirements may be more restrictive than industry or regulatory requirements.  This particular product being a fielded product has a Product Specification, this document defines a level of accuracy of the radar altimeter that must be maintained post-modification.  Table \ref{Tab:1a} shows the expected accuracy of the radar altimeter.

\begin{table}[htbp]
\caption{TSO-C87 Accuracy Requirements (Reproduced from: TSO-C87, 1966)}
\begin{center}
\begin{tabular}{|c|c|}
\hline
\textbf{Altitude $AL$ (ft.)}& \textbf{Rad Alt Output Accuracy}\\
\hline
$3ft \leq AL < 100ft$ & $\pm3ft$\\
\hline
$100ft \leq AL < 500ft$  & $\pm3$\%\\
\hline
$AL \geq 500ft$ & $\pm5$\%\\
\hline
\end{tabular}
\label{Tab:1a}
\end{center}
\end{table}


Consideration must be given to non-regulatory requirements that are sometimes imposed by a customer who is typically the aircraft manufacturer or operator, the radar altimeter in this case study is not subject to such customer requirements and will not be discussed further. 

\textbf{Compliance Strategy and MOC}
The purpose of Compliance is to associate a given product with a performance standard.  In this study the dominant standard is TSO-C87.  Due to the harsh environment that avionics equipment must operate in, there is a need to prove satisfactory operation when subjected to elements such as vibration, shock, temperature gradients, humidity, fluids, sand, dust and EMI \cite{RTCAInc1974MinimumAltimeters}. 
A sample of possible Affected Regulations/Requirements/Standards is presented in Table \ref{Tab:2a} along with the proposed MOCs.  The MOC is the technique that the RALT manufacturer will use to Show Compliance to the standards.  The table further shows that not only are different MOCs used but that they may be used in combination.

\begin{table}[htbp]
\caption{Sample Compliance and MOC}
\begin{center}
\begin{tabular}
{|p{0.1\textwidth}|p{0.15\textwidth}|p{0.1\textwidth}|}
\hline
\textbf{TSO Standard\/Other}& \textbf{MOPS}& \textbf{MOC} \\
\hline
TSO-C87 & AMOC & Tests, Analysis\\
\hline
Non-Regulatory &Product Specification& Tests\\
\hline
\end{tabular}
\label{Tab:2a}
\end{center}
\end{table}


\textbf{Verification Methods}
The two verification methods chosen for discussion are Analysis and Tests.  

Analysis as a MOC refers to the use of design documentation such as engineering drawings, product specifications and system descriptions to prove that a given product meets the intent or is Compliant with a particular requirement.  An engineering drawing showing the updated architecture of the RALT including the bypass filter, can be used as partial evidence to Show Compliance to the requirement that a filter must be installed to protect against interference.  Similarly, a vendor-provided product specification for the filter itself can be used to show how its performance is adequate to protect against the specific interference frequencies by referencing the Stopbands.  It is common practice to use product specification documentation to Show Compliance with environmental conditions, as the vendor will typically include details of the environment in that their product is designed to work in.

Tests are an actual demonstration of the performance of the product being evaluated during scripted scenarios.  There is a setup procedure for the test, there are the steps to be performed, and then there is the test outcome. Tests as a MOC are further reduced to Laboratory Tests and Flight Tests; aircraft ground testing is regarded as Flight Tests. 

\textbf{Return to Service}
A plan needs to be constructed as to how the modified RALTs will re-enter service. The typical means is a Service Letter (SL) for less critical items or a Service Bulletin (SB) to communicate more critical aircraft and equipment updates to the operator.  Given that the presence of 5G interference has generated safety concerns, it is appropriate to issue a Service Bulletin.  A date is defined that the SB needs to be complied with or else the grounding of the aircraft may occur. 
The SB is transmitted to all owners and operators of the affected aircraft on a serialization basis.  The SB contains a history of the problem and a set of instructions for remedying the concern.  The SB for this issue will detail the removal of the legacy and non-compliant RALT and detailed instructions for installation and checkout of the modified RALT. 

Though the modified RALT will have a TSO and, therefore FAA approval, there is a different certification path to install the RALT and integrate it with other avionics components for use in flight.  Another obstacle to be overcome is that the revised RALT will have a part number different to that which it is replacing and therefore will void the aircraft’s Type Certification (assumes no approved substitution via Illustrated Parts Catalog (IPC)).  A common solution to this is to apply for a Supplemental Type Certificate (STC) which if approved, modifies the original aircraft Type Certificate.

\section{Re-Design Assessment and Verification}\label{verification}
\textbf{Analysis:} Analysis has been selected as one of the MOC for Showing Compliance, refer to Table \ref{Tab:2a}.  It is expected that the RALT performs satisfactorily when exposed to intentional and spurious emissions from adjacent authorized spectrum users.  The specification for a suitable filter will reveal that, attenuation does not occur between 4000 MHz and 4600 MHz hence creating a Bandpass consistent with the required range for a RALT of $4.2 GHz-4.4 GHz$ ($4200 MHz–4400 MHz$).  The Stopbands are to be defined below and above the operational rad alt ranges. 

\textbf{Tests:}
Laboratory tests are planned as a MOC to Show Compliance.  Indeed, laboratory tests provide a means to simulate direct exposure to a 5G environment with the test engineer having the ability to vary the intensity of the interference in a controlled manner.  The modified RALT is sent an input signal via the Rx RF port to simulate a known altitude.  This attitude is confirmed via a recording of the outputs of the RALT.  The test is repeated. However, the input signal is distorted by frequencies representative of 5G and the RALT output of computed altitude is recorded. 

Flight tests provide the opportunity to expose the units under test to a real-world environment and allow the equipment to be used in a manner consistent with the intended application.   

Due to the potentially hazardous nature of flight test activities, a Safety of Flight Letter is required of the Engineering team responsible for the modification of the rad alt.  The letter provides evidence that the modified unit has been subjected to laboratory-based testing and does not present a hazard of fire, explosion or EMI to the basic aircraft systems when powered on.  The Safety of Flight Letter also lists the Serial Number of the RALT to be tested as this is validated at the time of installation in the test aircraft. 

As the modified radar altimeter is not TSO’d at the time of test execution an alternate means of approval needs to be used to install and fly the aircraft with unapproved equipment.  The aircraft is placed in an Experimental Category for purposes of Research and Development and at the time of Certification in Show Compliance using a Conformity process.\cite{DepartmentofTransportationFederalAviationAdministration2017AirworthinessAircraft}.  This is a special aircraft Category issued by the FAA for the purpose of flight testing. 

Per FAA guidance the rad alt system with an added filter should perform its function of altitude measurement with similar accuracy of the un-modified rad alt.  A dual radar altimeter installation is recommended to enable this comparison.  Performance demonstrations of the two rad alts will ensure performance consistent with the legacy product.  For this type of requirement which is performance-based, Flight Tests are an appropriate MOC. 

The radar altimeter has both transmit and receive components.  Given that the flight test will require this bi-directional signal propagation over a geographic expanse, prior approval from the FCC is required.  An application must be filed in advance. 

For new products the FAA must approve or may delegate their approval to conduct the flight test campaign.  The FAA also reserves the right to observe any Certification flight test even if approval is delegated.  The approval follows their review of the Flight Test Plan. 

The Flight Test Plan is the formal means of communicating to the FAA the details of the flight test campaign.

\textbf{Supplemental Type Certificate:} The RALT manufacturer provides the aircraft manufacturer with technical and certification details to show that the replacement RALT meets or exceeds the performance of the previous RALT and therefore does not compromise the aircraft.  Installation drawings for the RALT need to be updated. However, they are contained in a separate engineering package for the STC, this maintains a clear path of independence from the original Type Certificate.  Compliance and Verification activities are also performed at the aircraft level to prove a successful integration of the modified RALT into the aircraft.  This standalone STC package, once approved by the Regulatory Authority, then allows the use of the modified RALT in the aircraft.  The STC process is critical in the return to service of the aircraft as it serves as the final level of approval for flight with equipment that differs from that of the original aircraft Type Design.

At the time of authoring this document, the lab testing to support the modified radar altimeter had not been completed. Only preliminary flight tests had been completed.  However, the test results are suggestive that the modified radar altimeter performs in a manner consistent with that of an un-modified unit.  


\section{Dicussion}\label{discussion}
Beyond the current results in this paper, the ongoing work includes an increased focus on a radar altimeter design that is less susceptible to EMI and system testing that exceeds the minimum requirements for certification. Specifically, robustness testing should target exposing the system to corner cases that are atypical for usual operation and regulatory impact on updating the existing Minimum Operating Performance Standards (MOPS) DO-155 to address the risk of 5G interference specifically.

There is much work to be accomplished in the near future to address radar altimeters and concerns about 5G. The regulations must be revised to reflect the new performance requirements and the minimum standards that the radar altimeters must be produced to. The RTCA states, “In all cases (TSO-C87, DO-155, and ED-30), there have been no specific requirements regarding interference susceptibility or receiver masks. The latest update to requirements was 1980 – what did the RF spectrum look like then?” \cite{Robin2020InterferenceSystems}. Given that radar altimeters are inherently wideband systems, they are potentially more susceptible to signal blocking than other types of receivers \cite{Robin2020InterferenceSystems}. As a result, the aviation industry should continue to expand its research on using passive devices, such as filters, that protect the radar altimeter transceiver. Other alternatives can be considered. In Japan, placement of 5G base stations is to be avoided within 200m from the approach path of aircraft. According to the Japanese scientists, this mitigation is effective in avoiding the blockage of radar altimeter signals \cite{Corfield2022JapanIt_new}. Japan has also experimented with 5G antenna configurations and demonstrated how beam formation can impact interference on radar altimeters \cite{ICAO2021FrequencyCampaign_new}. Yet another alternative is increased band separation between 5G and radar altimeters. This is the approach employed in Europe. European 5G utilizes 3.4 to 3.8 GHz, whereas US deployments of 5G use 3.7 to 3.98 GHz, thus reducing the margin from the 4.2 GHz lower limit of radar altimeters \cite{Leslie2022USSafety}. At present, most 5G efforts assume power levels typical for 5G systems in the US. Additional research, however, is needed to ensure global compatibility as 5G deployment in Europe, for example, is subject to higher power levels, up to 1.5 times higher. Indeed, it is expected that domestically, the power levels of 5G will also increase eventually. The radar altimeter and 5G conflict highlight another area of concern, that of complacency. The radar altimeter technology has not advanced from a security or threat perspective, yet its RF environment is becoming increasingly threatened. A review should be conducted of other known critical avionics systems that similarly have benefited from quiet RF environments and therefore do not have protections designed into them.


The FAA's recommended solution of an RF filter is effective against 5G interference. 
All development and production product names and part numbers within this study have been altered to be disassociated from any manufacturers’ product line.  Any similarities are unintentional. 

\section{Conclusions and Future Research}\label{conclusion}

This study presents the methods that may be used by a radar altimeter manufacturer to increase the robustness of a hypothetical RALT against 5G interference. The incomplete test results of this study suggest that the addition of a bandpass filter intended to limit frequencies other than the 4.2 to 4.4 GHz does not negatively impact the basic radar altimeter function. Though the laboratory testing is incomplete at this time, preliminary test results are commensurate with those observed in flight tests. As the unmodified radar altimeter is an approved unit with a TSO, it is proposed in this study as a control. The accuracy of the TSO'd unit meets those of the regulatory requirements in Table \ref{Tab:2a}.

There is a history of aircraft accidents resulting from radar altimeter anomalies. Turkish Airlines Flight 1951 was caused primarily by the aircraft's automated reaction triggered by a faulty radio altimeter. This accident resulted in fatalities. On December 25th, 2012, in Kazakhstan, an Antonov 72 crashed, killing all 20 onboard. Following an autopilot failure, the captain "decided to fly the plane manually. Two minutes and 40 seconds after takeoff, the radio altimeter also failed. The flight was continued using barometric altimeters… there was a momentary failure of these altimeters as well…" \cite{AviationSafeyNetwork2013AltimeterKazakhstan.}. The airplane collided with terrain 21km short of the runway and broke apart.

The FAA has also been clear that there is a need, "to establish a timeline for retrofitting or replacing radar altimeters in US airliners that are affected by 5G C-band signals…" \cite{Vigliarolo2022FAAGo_new}. 5G potential interference will not be ignored, and there are regulatory consequences. RALT manufacturers will need to improve their current product line to make them less 5G susceptible. This study provides insights into a strategic solution for radar altimeters for actual flights for multi-stage airborne interference risk mitigation in 5G and future G networks.

\bibliographystyle{IEEEtran} 
\bibliography{references,references_ying}


\begin {IEEEbiography}[{\includegraphics[width=1in,height=1.25in,clip,keepaspectratio]{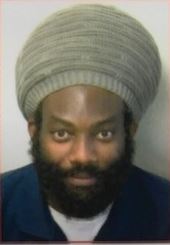}}]{Jarret Rock} (Member, IEEE) received the B.Sc. degree in Aeronautical Science at Florida Institute of Technology, M.Sc. degree in Aviation Human Factors from Florida Institute of Technology and M.S. degree in Systems Engineering from Stevens Institute of Technology. He is an industry professional Test Pilot and Systems Engineer with specialties in avionics certification, systems integration and flight test. His research areas include cybersecurity, aircraft safety, aircraft autonomy, avionics systems and systems engineering. 
\end{IEEEbiography} 
\begin {IEEEbiography}[{\includegraphics[width=1in,height=1.25in,clip,keepaspectratio]{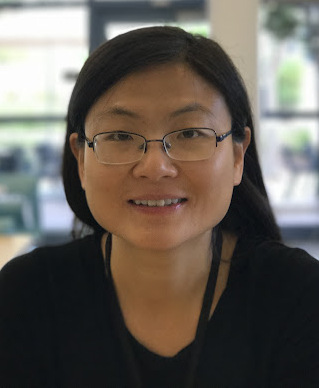}}]{Ying Wang} (Member, IEEE) received the B.E. degree in information engineering at Beijing University of Posts and Telecommunications, M.S. degree in electrical engineering from University of Cincinnati and the Ph.D. degree in electrical engineering from Virginia Polytechnic Institute and State University. She is an associate professor in the School of System and Enterprises at Stevens Institute of Technology. Her research areas include cybersecurity, wireless AI, edge computing, health informatics, and software engineering. 
\end{IEEEbiography}

\end{document}